\documentclass[12pt]{article}
\usepackage{graphicx} 
\usepackage{amsmath}
\usepackage{amssymb}
\usepackage{xcolor}
\usepackage[hyperindex=true,
          pdfstartview=FitH,
          bookmarksnumbered=true,
          bookmarksopen=true,
          citecolor=blue,
          linkcolor=blue,
          colorlinks=true,
          urlcolor=purple,
          unicode]{hyperref}

\usepackage[top=2.0cm,bottom=2.0cm,left=2.54cm,right=2.54cm]{geometry}
\usepackage{cite}

\usepackage{xcolor}

\title{Covariant linear response theory for a photon gas in curved spacetime}
\author{
Jianan Wang,${}^1$~
Long Cui,${}^1$~
and Bin Wu ${}^{1,2,3,4}$\thanks{{\em email}: \href{mailto:binwu@nwu.edu.cn}{binwu@nwu.edu.cn}}\\
\small ${}^1$School of Physics, Northwest University, Xi'an\ 710127, China\\
\small ${}^{2}$Shaanxi Key Laboratory for Theoretical Physics Frontiers, Xi'an 710127, China\\
\small ${}^{3}$Peng Huanwu Center for Fundamental Theory, Xi'an 710127, China\\
\small ${}^{4}$Fundamental Discipline Research Center for Quantum Science and Technology of\\
\small Shaanxi Province, Xi'an 710127, China
}

\date{}
\allowdisplaybreaks[4]
\begin{document}

\maketitle
\begin{abstract}
To overcome the failure of the conventional non-relativistic description of photon statistics in strong gravitational fields and for non-comoving observers, we develop a covariant theory for a photon gas in curved spacetime, starting from the relativistic Boltzmann equation and employing the relaxation time approximation. The transport coefficients are computed up to second order in the relaxation time. The results reveal novel phenomena absent in flat space. At second order, the gravitational field renders the transport coefficients tensor-valued and induces transverse particle and energy fluxes, analogous to the Hall effect in electromagnetism. Using the first-order results, we derive the relativistic generalizations of Fick's and Fourier's laws and obtain the radiation diffusion equation for spherically symmetric accretion disks around Schwarzschild black holes. This provides a covariant description of radiative transport in strong gravitational fields for high-energy astrophysical applications.
\end{abstract}

\newpage

\section{Introduction}

Planck's blackbody radiation formula and the Stefan-Boltzmann law~\cite{Greenberger_2007, Nemova_2022} are cornerstones of equilibrium statistical physics, establishing the relation between photon energy density and temperature in a thermal bath. However, these expressions are fundamentally derived under the assumption that the observer is comoving with the system. In the presence of relative motion between the observer and the system, the conventional non-relativistic description of photon statistics becomes observer-dependent and loses its covariant meaning~\cite{Hsu_1992}.

This limitation is not merely a theoretical subtlety. In realistic astrophysical environments such as stellar interiors, compact object atmospheres, and black hole accretion disks, matter undergoes relativistic bulk motion and resides in strong gravitational fields. In these regimes, the non-equilibrium transport of radiation is inevitably coupled to spacetime curvature, and a consistent description requires a fully relativistic, observer-independent formulation of photon kinetics.

Early advances toward a relativistic statistical theory were made by J\"{u}ttner~\cite{Juttner_1911}, who derived the equilibrium single-particle distribution function (1PDF) for relativistic gases. Subsequently, Tolman and Ehrenfest~\cite{Tolman1930,TolmanEhrenfest1930} established a fundamental condition for thermal equilibrium in a static gravitational field,
\begin{equation}
\vec{\nabla} T = -T \vec{\nabla}\phi ,\label{Tolman and Ehrenfest}
\end{equation}
which implies that a temperature gradient is necessary to counteract the gravitational redshift. This relation already indicates that the simple Fourier law cannot hold covariantly in curved spacetime, since equilibrium would then require a vanishing temperature gradient, contradicting Eq.~(\ref{Tolman and Ehrenfest}).

More recently, Sarbach and collaborators~\cite{Sarbach_2014, Sarbach_2014_2, Sarbach_2017, Acuna_2022} developed a rigorous geometric framework for relativistic kinetic theory on tangent and cotangent bundles, providing a solid foundation for covariant statistical mechanics of massive particles. Related studies~\cite{Rioseco_2017, Cieslik:2020ibk, Cai:2022fdu, Liao:2022lza, Shapiro:2023gpe, Li:2023qtd, liu2026accretionplasmavlasovgas} have applied to high-energy astrophysical systems such as black hole accretion disks, though without incorporating particle interactions. This is partly due to the intrinsic difficulties of gravitational thermal effects in non-equilibrium regimes, where both observational constraints and theoretical complexity remain challenging~\cite{Kijowski_1997, Rovelli_2011, Xia_2024}. In the literature~\cite{Hao_2024, Cui_2025_2}, the authors constructed a covariant description for massive particle systems using the Boltzmann equation under the relaxation time approximation. In particular, the transport coefficients were systematically computed up to second order in the relaxation time, revealing that spacetime curvature induces novel non-diagonal transport phenomena absent in the flat space limit~\cite{Cui_2025}.

However, a crucial distinction must be made between massive particles and massless photons. For massive particles, the mass-shell condition \(g_{\mu\nu}p^\mu p^\nu = -m^2\) permits a chemical potential associated with particle number conservation. Photons, by contrast, obey the null condition \(g_{\mu\nu}p^\mu p^\nu = 0\) and carry no conserved particle number (hence zero chemical potential). Their lack of a mass scale makes the kinetic theory structurally different from that of massive particles~\cite{Cai:2026fym}. And in the photon case the phase-space measure, the equilibrium distribution, the collision operator, and the coupling between dissipative fluxes all differ fundamentally. These differences are not merely quantitative adjustments, so a self-contained derivation for photons is imperative.

The importance of such a derivation extends beyond formal theory. In high-energy astrophysical environments, such as the interior of massive stars, the photospheres of neutron stars, and the accretion disks around black holes, radiation is the dominant mechanism of energy transport and cooling. The photons that carry this energy propagate through strongly curved spacetimes and interact frequently with the ambient plasma via emission, absorption, and Compton scattering. A quantitative understanding of these processes, from the dynamics of core-collapse supernovae to the stability of accretion flows, relies crucially on a covariant theory of radiative transport that correctly accounts for both non-equilibrium statistics and gravitational effects. Previously, Udey and Israel~\cite{Udey1982} studied general-relativistic radiative transfer and radiation hydrodynamics within the 14-moment approximation, and evaluated the transport coefficients and the corresponding equations of motion.

In contrast, here we construct a covariant linear-response theory for a photon gas based on an iterative expansion in the relaxation time, without truncating in moment hierarchy. Starting from the relativistic Boltzmann equation within the relaxation time approximation (RTA), we compute the photon transport coefficients up to second order in the relaxation time and derive the corresponding macroscopic transport laws. This approach yields several novel features in the transport phenomena, including tensor-valued transport coefficients and transverse transport induced by the gravitational field. The resulting radiation diffusion equation is then applied to spherically symmetric accretion disks around Schwarzschild black holes. We adopt natural units (\(c = k_B = 1\)) and consider a \((d+1)\)-dimensional spacetime with metric signature \((-, +, \dots, +)\). Greek indices \(\mu, \nu, \dots\) denote coordinate components, while early Latin indices \(a, b, \dots\) denote orthonormal frame components.

\section{System Constituted by Photon Gas in Equilibrium}
The system under consideration consists of a background spacetime $g_{\mu\nu}$, a photon gas, and a relativistic material medium in local thermodynamic equilibrium. The medium acts as a heat reservoir, while the photon gas is an open subsystem embedded in it. Since both are relativistic, their macroscopic currents and energy-momentum tensors are defined in the same way from their respective distribution functions. Unless otherwise stated, the single-particle distribution function (1PDF) \(f\), the energy-momentum tensor \(T^{\mu\nu}\), and all other physical quantities refer exclusively to the photon gas, while the corresponding quantities for the material medium are denoted by a subscript \(r\), and the total energy-momentum tensor is $T^{\mu\nu}_{\rm tot}=T^{\mu\nu}_{r}+T^{\mu\nu}$. Throughout this work, the four-velocity $U^{\mu}$ and the temperature \(T\) are those of the photon gas. They define the frame in which the photon distribution function takes its local-equilibrium form, and they are the variables that enter the hydrodynamic description of the photon gas developed below. In equilibrium the photon gas and the medium share the same temperature and four-velocity fields, because the vanishing of the collision integral ties the photon distribution to that of the medium. Away from equilibrium the two may differ, and the medium may then carry its own four-velocity and temperature. The medium is not described hydrodynamically here, its entire influence on the photon gas, including the associated relaxation and thermalization processes, is compressed into the relaxation time $\tau$ introduced later.

For a relativistic system, the macroscopic particle current \(N^{\mu}\) and energy-momentum tensor \(T^{\mu\nu}\) can be expressed as moments of the 1PDF. We begin by defining these integral relations, which are valid for both equilibrium and non-equilibrium states \cite{Sarbach_2014,Sarbach_2014_2,Sarbach_2017}:
\begin{equation}
    N^{\mu} = \int\varpi p^{\mu} f, \quad T^{\mu\nu} = \int\varpi p^{\mu} p^{\nu} f.
    \label{int}
\end{equation}
Here \(p^{\mu}\) denotes the particle momentum, \(f\) is the 1PDF, and \(\varpi=\frac{\sqrt{|\det(g_{\mu\nu})|}}{|p_0|}\mathrm{d}^dp\) is the covariant volume element in momentum space, which ensures the integrals are coordinate-invariant.

The evolution of \(f\) is governed by the relativistic Boltzmann equation \cite{1980Relativistic}
\begin{equation}
    \mathcal{L}[f] = \mathcal{C}[f, f_r],
    \label{Boltzmann equation}
\end{equation}
where \(\mathcal{L}\) is the Liouville vector field for free particles,
\begin{equation}
    \mathcal{L} = p^{\mu} \frac{\partial}{\partial x^{\mu}} - {\Gamma^{\mu}}_{\nu\lambda} p^{\nu} p^{\lambda} \frac{\partial}{\partial p^{\mu}},
\end{equation}
and \(\mathcal{C}\) is the collision integral. Treating the massive particles (material medium) as a thermal reservoir, we regard the photon gas as an open subsystem, so that \(\mathcal{C}[f, f_r]\) describes photon-matter interactions such as emission, absorption, and Compton scattering. 

In equilibrium, the collision integral vanishes \cite{1980Relativistic}
\begin{equation}
    \mathcal{C}[f_{\rm eq}, f_r] = 0.
\end{equation}
For bosons such as photons, the solution is the Bose-Einstein distribution with zero chemical potential:
\begin{equation}
    f_{\rm eq} = \frac{\mathfrak{g}_{0}}{h^{d}} \frac{1}{\exp(-\mathcal{B}_{\mu} p^{\mu}) - 1},
    \label{equilibrium single-particle distribution function}
\end{equation}
where \(h\) is Planck's constant, \(\mathfrak{g}_{0}\) is the degeneracy factor (\(\mathfrak{g}_{0}=d-1\) for photons), and \(\mathcal{B}^{\mu}=\beta U^{\mu}\), with \(U^{\mu}\) the fluid four-velocity and \(\beta=1/T\) the inverse temperature.

Substituting Eq.~(\ref{equilibrium single-particle distribution function}) into the Boltzmann equation yields
\begin{equation}
    \mathcal{L}[f_{\rm eq}]= p^{\mu}p^{\nu}\nabla_{(\mu}\mathcal{B}_{\nu)}\frac{\partial f_{eq}}{\partial \xi}=0, \quad \xi=-\mathcal{B}_{\mu}p^{\mu}.
    \label{photon system in equilibrium}
\end{equation}
Since \(p^{\mu}\) is arbitrary on the mass shell, Eq.~(\ref{photon system in equilibrium}) is satisfied if and only if \(\mathcal{B}^{\mu}\) is a Killing vector field:
\begin{equation}
    \nabla_{(\mu} \mathcal{B}_{\nu)}=0.
    \label{detailed balance condition}
\end{equation}
This is the condition for \emph{global} thermodynamic equilibrium. In the next section, we will relax this condition and adopt the \emph{local} equilibrium assumption, allowing \(\beta\) and \(U^\mu\) to vary slowly in spacetime.

To evaluate the equilibrium integrals explicitly, we introduce a local orthonormal frame \(\{(e^{a})_{\mu}\}\) satisfying \(\eta_{ab}=g_{\mu\nu}(e_{a})^{\mu}(e_{b})^{\nu}\), and choose it such that the time-like basis vector coincides with the fluid four-velocity, i.e., \(U^{\mu}=(e_{0})^{\mu}\). In this frame, the photon momentum satisfies
\begin{equation}
    \eta_{ab}p^a p^b = -(p^0)^2 + \sum_{i=1}^d (p^i)^2 = 0,
\end{equation}
so that \(p^a\) can be parametrized as
\begin{equation}
    p^{a}=(p^0,p^0 n^i),\label{momentum}
\end{equation}
where \(n^{i}\) is the unit normal vector on the \((d-1)\)-sphere, satisfying \(n_i n^i=1\). 

The Cartesian momentum volume element \(\mathrm{d}^d p\) can then be expressed in spherical coordinates as
\begin{equation}
    \mathrm{d}^d p = (p^0)^{d-1} \mathrm{d}p^0 \, \mathrm{d}\Omega_{d-1},
\end{equation}
where \(\mathrm{d}\Omega_{d-1}\) is the solid angle on the \((d-1)\)-sphere. Substituting this and \( \sqrt{|\det(\eta_{ab})|}=1\) into the covariant measure \(\varpi = \frac{\sqrt{|\det(g_{\mu\nu})|}}{|p_0|}\mathrm{d}^d p\) yields
\begin{equation}
    \varpi=(p^0)^{d-2}\,\mathrm{d}p^0\,\mathrm{d}\Omega_{d-1}.
    \label{volume element}
\end{equation}

In the following integrals, we use the standard formula
\begin{equation}
    \int_0^\infty \frac{z^{s-1}}{e^z - 1} \mathrm{d}z = \Gamma(s)\zeta(s),
\end{equation}
where \(\Gamma(s)\) is the Gamma function and \(\zeta(s)=\sum_{k=1}^{\infty}k^{-s}\) is the Riemann zeta function, and the angular identity
\begin{equation}
    \int \mathrm{d}\Omega_{d-1}\, n^i n^j = \frac{1}{d}\delta^{ij}\Omega_{d-1},
\end{equation}
with \(\Omega_{d-1}=2\pi^{d/2}/\Gamma(d/2)\) the area of the \((d-1)\)-sphere.

Substitute Eq.~(\ref{equilibrium single-particle distribution function}) into Eq.~(\ref{int}), and use the momentum decomposition Eq.~(\ref{momentum}) in the orthonormal frame. The angular integrals then eliminate the linear terms in \(n^i\) because the contributions from opposite directions cancel by spherical symmetry, while the quadratic terms are reduced using the identity above. We obtain the equilibrium particle current
\begin{equation}
    \begin{split}
        N^{\mu}_{\rm eq}&=\frac{\mathfrak{g}_{0}}{h^{d}}\int \mathrm{d}p^0\,\mathrm{d}\Omega_{d-1}
        \frac{(p^0)^{d-1}}{\exp(\beta p^{0}) - 1}
        \left( U^{\mu} + n^i (e_i)^\mu \right) \\
        &= \frac{\mathfrak{g}_{0}}{h^{d}}\int \mathrm{d}p^0\,\mathrm{d}\Omega_{d-1}
        \frac{(p^0)^{d-1}}{\exp(\beta p^{0}) - 1}\,U^{\mu} \\
        &=\frac{\mathfrak{g}_{0}}{h^{d}}\Omega_{d-1}\Gamma(d)\zeta(d)T^{d}U^{\mu},
    \end{split}
\end{equation}
and the equilibrium energy-momentum tensor,
\begin{equation}
    \begin{split}
        T^{\mu\nu}_{\rm eq}&=\frac{\mathfrak{g}_{0}}{h^{d}}\int \mathrm{d}p^0\,\mathrm{d}\Omega_{d-1}
        \frac{(p^0)^{d}}{\exp(\beta p^{0}) - 1}
        \left( U^{\mu}U^{\nu} + U^{\mu}n^j(e_j)^\nu + n^i(e_i)^\mu U^\nu + n^i n^j (e_i)^\mu (e_j)^\nu \right) \\
        &= \frac{\mathfrak{g}_{0}}{h^{d}}\int \mathrm{d}p^0\,\mathrm{d}\Omega_{d-1}
        \frac{(p^0)^{d}}{\exp(\beta p^{0}) - 1}
        \left( U^{\mu}U^{\nu} + \frac{1}{d}\Delta^{\mu\nu} \right) \\
        &=\frac{\mathfrak{g}_{0}}{h^{d}}\Omega_{d-1}\Gamma(d+1)\zeta(d+1)T^{d+1}
        \left(U^{\mu}U^{\nu}+\frac{1}{d}\Delta^{\mu\nu}\right).
    \end{split}
\end{equation}
where \(\Delta^{\mu\nu} = \delta^{ij}(e_i)^\mu(e_j)^\nu\) is the projection tensor introduced via the orthonormal frame.

These integral results can now be compared with the macroscopic decomposition of the particle current and energy-momentum tensor for a perfect fluid in equilibrium:
\begin{equation}
    N^{\mu} = n U^{\mu}, \quad T^{\mu\nu} = \epsilon U^{\mu} U^{\nu} + P \Delta^{\mu\nu}.
    \label{perfect}
\end{equation}
Here \(\Delta_{\mu\nu} = g_{\mu\nu} + U_{\mu}U_{\nu}\) is the projection tensor onto the spatial hypersurface orthogonal to \(U^\mu\), which is equivalent to the expression \(\Delta^{\mu\nu} = \delta^{ij}(e_i)^\mu(e_j)^\nu\) used in the orthonormal frame, and \(n\), \(\epsilon\), \(P\) are the particle number density, energy density, and isotropic pressure measured by an observer comoving with the fluid (i.e., an observer whose four-velocity coincides with \(U^\mu\)). Comparing the integral results with this decomposition yields
\begin{equation}
    \begin{split}
        n_{\rm eq}&=-U_{\mu}N^{\mu}_{\rm eq}\\
         &=\frac{\mathfrak{g}_{0}}{h^{d}}\Omega_{d-1}\Gamma(d)\zeta(d)T^{d},\\
        \epsilon_{\rm eq}&=U_{\mu}U_{\nu}T^{\mu\nu}_{\rm eq}\\
        &=\frac{\mathfrak{g}_{0}}{h^{d}}\Omega_{d-1}\Gamma(d+1)\zeta(d+1)T^{d+1},\\
        P_{\rm eq}&=\frac{1}{d}\Delta_{\mu\nu}T^{\mu\nu}_{\rm eq}\\
        &=\frac{\mathfrak{g}_{0}}{h^{d}}\Omega_{d-1}\Gamma(d)\zeta(d+1)T^{d+1}.
    \end{split}
    \label{functions of temperature}
\end{equation}
These are the equilibrium thermodynamic relations for a photon gas in \(d+1\)-dimensional curved spacetime, evaluated in the comoving frame.

For \(d=3\), the energy density reproduces the blackbody energy density–temperature relation. The above results confirm that the non-relativistic formulation applies only to measurements performed by an observer comoving with the fluid. To extend the description to non-equilibrium states, the next section develops a linear-response theory in which deviations from local equilibrium are treated as small perturbations driven by thermodynamic forces.

\section{Linear Response of the Photon System}
Before proceeding, we clarify the expansion scheme employed in this work. Two distinct expansions are involved:
\begin{enumerate}
    \item \emph{Expansion in the relaxation time:} The RTA iterative solution yields corrections to the distribution function at successive powers of the relaxation time \(\tau\). We retain terms up to second order in \(\tau\), as this is the lowest order at which the coupling between spacetime curvature and thermodynamic forces appears.
    
    \item \emph{Gradient expansion:} This refers to the expansion in powers of gradients of the thermodynamic quantities (e.g., temperature, fluid velocity). In computing the transport coefficients, we retain contributions from all gradient orders, provided they are linear in the thermodynamic forces and  all nonlinear products are discarded.
\end{enumerate}
The resulting linear-gradient fluxes presented later receive contributions from both the first-order and second-order RTA corrections.

In the previous chapter, the distribution Eq.~(\ref{equilibrium single-particle distribution function}) of a photon gas was fully characterized by the fluid four-velocity \(U^{\mu}\) and the parameter \(\beta\). Out of equilibrium, the distribution function depends on additional degrees of freedom beyond these local parameters, namely the thermodynamic forces that drive the system back to equilibrium. To extend the description to systems with slight deviation from equilibrium, we employ the local equilibrium assumption. The zeroth-order term of the 1PDF is given by the same functional form as in equilibrium,
\begin{equation}
    f^{(0)} = \frac{\mathfrak{g}_0}{h^{d}} \frac{1}{\exp(-\mathcal{B}_{\mu} p^{\mu}) - 1},
\end{equation}
but now \(\mathcal{B}^{\mu} = \beta U^{\mu}\) is allowed to vary slowly in spacetime. The Killing condition is relaxed, so that the local equilibrium distribution no longer makes the Liouville operator vanish \(\mathcal{L}[f^{(0)}] \neq 0\).

The task is to determine the complete 1PDF by combining its zeroth-order term with the Boltzmann equation. However, since the Boltzmann equation is an integro-differential equation, obtaining an analytical solution directly is nearly impossible, and even numerical solutions are very challenging. Fortunately, for small deviations from equilibrium, the collision integral can be replaced by a simple relaxation-time term, which brings the Boltzmann equation into a tractable form \cite{Anderson_1974}:
\begin{equation}
    \mathcal{L}[f]=-\frac{\varepsilon}{\tau}(f-f^{(0)}),
    \label{relaxation time approximation}
\end{equation}
where \(\varepsilon = -p^{\mu}U_{\mu}\) is the single-photon energy in the fluid frame, and \(\tau\) is the relaxation time \footnote{In Minkowski spacetime with a homogeneous fluid, Eq.~(\ref{relaxation time approximation}) reduces to \(df/dt = -(f-f^{(0)})/\tau\), whose solution is \(f(t) = f^{(0)} + (f(0)-f^{(0)})e^{-t/\tau}\). Thus \(\tau\) characterizes the timescale for the system to return to equilibrium.}, which is determined by the properties of the fluid and the 1PDF of the massive particles $f_r$. The relaxation time therefore encodes the entire influence of the material medium on the photon gas, including the four-velocity and temperature of the medium, which may differ from \(U^{\mu}\) and \(T\) away from equilibrium.

The arbitrary-order term of the 1PDF can be obtained as
\begin{equation}
    f^{(n)} = \left(-\frac{\tau}{\varepsilon}\mathcal{L}\right)^n[f^{(0)}].
    \label{arbitrary-order term}
\end{equation}
The explicit first-order term is then
\begin{equation}
    f^{(1)} = -\frac{\tau}{\varepsilon} \mathcal{L}[f^{(0)}]
    = -\frac{\tau}{\varepsilon} p^{\mu}p^{\nu}\nabla_{(\mu}\mathcal{B}_{\nu)}
    \frac{\partial f^{(0)}}{\partial \xi}, \quad \xi \equiv -\mathcal{B}_{\mu}p^{\mu}.
    \label{first-order}
\end{equation}
Thus \(f^{(1)}\) is proportional to \(\nabla_{(\mu}\mathcal{B}_{\nu)}\), which vanishes in equilibrium. When the system deviates from equilibrium, its non-vanishing value drives the system back to equilibrium, and hence it is identified as the covariant thermodynamic force.

A photon system in a non-equilibrium state is no longer in global equilibrium (i.e., the Killing condition is not satisfied, so \(\nabla_{(\mu}\mathcal{B}_{\nu)}\neq 0\)). To separate the various mechanisms that drive the system back to equilibrium, we decompose \(\nabla_{(\mu}\mathcal{B}_{\nu)}\) into irreducible components with respect to the fluid four-velocity. Starting from \(\mathcal{B}_\mu = \beta U_\mu\), its symmetrized gradient can be expressed as
\begin{equation}
    \nabla_{(\mu}\mathcal{B}_{\nu)} = U_{(\nu}\nabla_{\mu)}\beta + \beta \nabla_{(\mu}U_{\nu)}.
\end{equation}

The symmetrized gradient of the fluid velocity is decomposed into temporal and spatial parts,
\begin{equation}
    \nabla_{(\mu}U_{\nu)}={\Delta_{(\mu}}^{\rho}{\Delta_{\nu)}}^{\sigma}\nabla_{\rho}U_{\sigma}-U_{(\mu}{\Delta_{\nu)}}^{\sigma}U^{\rho}\nabla_{\rho}U_{\sigma}
\end{equation}
Thus \(\nabla_{(\mu}\mathcal{B}_{\nu)} = U_{(\nu}\nabla_{\mu)}\beta+\beta\left({\Delta_{(\mu}}^{\rho}{\Delta_{\nu)}}^{\sigma}\nabla_{\rho}U_{\sigma}-U_{(\mu}{\Delta_{\nu)}}^{\sigma}U^{\rho}\nabla_{\rho}U_{\sigma}\right)\), the terms multiplied by \( U_{\nu} \) are decomposed into temporal and spatial parts,
\begin{equation}
    \nabla_\mu \beta-\beta{\Delta_{\mu}}^{\sigma}U^{\rho}\nabla_{\rho}U_{\sigma} = -U_{\mu} \dot{\beta} + \mathcal{D}_\mu\beta,
    \label{gradient beta}
\end{equation}
where
\begin{equation}
    \dot{\beta} \equiv U^\mu \nabla_\mu\beta,\qquad 
    \mathcal{D}_\mu\beta \equiv {\Delta_\mu}^\nu (\nabla_\nu\beta-\beta U^{\rho}\nabla_{\rho}U_{\sigma}).
\end{equation}
Thus \(U_{(\mu}\nabla_{\nu)}\beta -\beta U_{(\mu}{\Delta_{\nu)}}^{\sigma}U^{\rho}\nabla_{\rho}U_{\sigma}= -\dot{\beta}U_\mu U_\nu + U_{(\mu}\mathcal{D}_{\nu)}\beta\).

Similarly, the spatial part of symmetrized gradient of the fluid velocity is decomposed into its trace and traceless parts,
\begin{equation}
    {\Delta_{(\mu}}^{\rho}{\Delta_{\nu)}}^{\sigma}\nabla_{\rho}U_{\sigma} = -\frac{1}{\beta}\left(\frac{1}{d}\psi \Delta_{\mu\nu} + \psi_{\mu\nu}\right),
\end{equation}
where
\begin{equation}
    \psi \equiv -\beta \nabla_\mu U^\mu,\qquad 
    \psi_{\mu\nu} \equiv -\beta\left({\Delta_{(\mu}}^{\rho}{\Delta_{\nu)}}^{\sigma}\nabla_{\rho}U_{\sigma}-\frac{1}{d}\nabla_\rho U^\rho\Delta_{\mu\nu}\right).
\end{equation}

Combining these results, we obtain the decomposition
\begin{equation}
    \nabla_{(\mu}\mathcal{B}_{\nu)}
    = -\dot{\beta}U_{\mu}U_{\nu}
    + U_{(\mu}\mathcal{D}_{\nu)}\beta
    - \frac{1}{d}\psi \Delta_{\mu\nu}
    - \psi_{\mu\nu}.
    \label{decomposition}
\end{equation}
The four terms correspond, respectively, to the temporal variation of \(\beta\), its spatial gradient, the expansion, and the shear of the fluid. We refer to \(\dot{\beta}\) and \(\psi\) as the scalar thermodynamic forces, \(\mathcal{D}_{\mu}\beta\) as the vector thermodynamic force, and \(\psi_{\mu\nu}\) as the tensorial thermodynamic force. As will become clear below, the first-order correction \(f^{(1)}\) yields the standard transport coefficients in the absence of curvature effects.



To capture the curvature effects that appear at second order in the relaxation time, we introduce the gravitoelectromagnetic field
\begin{equation}
    E^{(G)}_{\mu}\equiv-U^{\rho}\nabla_{\rho}U_{\mu} ,\quad \quad B^{(G)}_{\mu\nu}\equiv{\Delta^{\rho}}_{\mu}{\Delta^{\sigma}}_{\nu}\nabla_{[\sigma} U_{\rho]}.
\end{equation}
The gravitoelectromagnetic field is manifestly observer-dependent \cite{Cui_2025}. Evaluating Eq. (\ref{arbitrary-order term}) explicitly using the Liouville operator, the second-order correction is obtained as
\begin{equation}
    \begin{split}
    f^{(2)} = &-\left(\frac{\tau}{\varepsilon}\right)^2\left[p^{\mu}p^{\nu}p^{\sigma}\nabla_{\sigma}(\nabla_{(\mu}\mathcal{B_{\nu)}})+\frac{1}{\varepsilon}(p^{\mu}p^{\nu}\nabla_{(\mu}U_{\nu)})(p^{\rho}p^{\sigma}\nabla_{(\rho}\mathcal{B_{\sigma)}})\right]\frac{\partial f^{(0)}}{\partial \xi}\\
    &-\left(\frac{\tau}{\varepsilon}\right)^2\left(p^{\mu}p^{\nu}\nabla_{(\mu}\mathcal{B}_{\nu)}\right)^{2}\frac{\partial^2 f^{(0)}}{\partial \xi^2}.
    \label{2t}
    \end{split}
\end{equation}
Unlike the first-order term, the second-order term contains derivatives of the thermodynamic forces (e.g., \(\nabla_{\mu}\dot{\beta}\), \({\Delta^{\rho}}_{\mu}{\Delta^{\sigma}}_{\nu}\nabla_{\sigma}(\mathcal{D}_{\rho}\beta)\)) and non-linear products of thermodynamic forces (e.g., \((\mathcal{D}\beta)^2\), \(\psi_{\mu\nu}\mathcal{D}_{\sigma}\beta\)). As we are working in the linear response regime, we discard these nonlinear products and retain only the linear part \(f^{(2)}_{\text{linear}}\), which is given in Appendix B.

On the other hand, a non-equilibrium fluid is no longer a perfect fluid. The particle current and energy-momentum tensor can be decomposed with respect to the fluid four-velocity \(U^\mu\) as
\begin{equation}
    N^{\mu} = (n^{(0)}+\delta n) U^{\mu}+j^{\mu},
    \label{non-equilibrium state1}
\end{equation}
\begin{equation}
    T^{\mu\nu} =( \epsilon^{(0)}+\delta\epsilon) U^{\mu} U^{\nu} +2q^{(\mu}U^{\nu)} +(P+\Pi) \Delta^{\mu\nu}+\Pi^{\mu\nu}.
    \label{non-equilibrium state2}
\end{equation}
Here \(n^{(0)}\) and \(\epsilon^{(0)}\) denote the local equilibrium values which take the same functional form as the equilibrium quantities in Eq.~(\ref{functions of temperature}), but are now allowed to vary slowly in spacetime. \(\delta n\), \(\delta\epsilon\) are the corresponding perturbations, \(j^\mu\) is the particle flux, \(q^\mu\) the energy flux, \(\Pi\) the dynamic pressure, and \(\Pi^{\mu\nu}\) the viscous stress tensor. These quantities are measured in the comoving frame. We refer to \(\delta n\), \(\delta\epsilon\), and \(\Pi\) as scalar dissipative quantities, and to \(j^\mu\), \(q^\mu\), and \(\Pi^{\mu\nu}\) as vector and tensorial dissipative terms, respectively.

Through the calculations in Appendixes~\ref{appendixA} and~\ref{appendixB}, we obtain the transport coefficients accurate to first and second order in the relaxation time. The equilibrium integrals evaluated in the previous section motivate the introduction of two temperature-dependent coefficients:
\begin{equation}
    \mathcal{C}_1\equiv\frac{\mathfrak{g}_{0}}{h^d}\Omega_{d-1}\Gamma(d+1)\zeta(d)T^{d+1},\quad\mathcal{C}_2\equiv\frac{\mathfrak{g}_{0}}{h^d}\Omega_{d-1}\Gamma(d+2)\zeta(d+1)T^{d+2}.
\end{equation}
Comparing the resulting kinetic expressions for the dissipative fluxes with their macroscopic decompositions in Eqs.~(\ref{non-equilibrium state1}) and~(\ref{non-equilibrium state2}), we obtain the following linear response relations, organized according to the tensor rank of the thermodynamic forces and dissipative fluxes. \indent(i) The scalar-scalar response (S-S) is
\begin{align}
&\left( \begin{matrix}
		\delta n \\
		\delta \epsilon \\
        \Pi
\end{matrix} \right)
= \tau 
\left( \begin{matrix}
	\mathcal{C}_1 & d^{-1}\mathcal{C}_1   \\
	\mathcal{C}_2 & d^{-1}\mathcal{C}_2 \\
	d^{-1}\mathcal{C}_2 & d^{-2}\mathcal{C}_2 \\
\end{matrix} \right)
\left( \begin{matrix}
		\dot{\beta} \\
        \psi
\end{matrix} \right). \label{S-S}
\end{align}
\indent (ii) The vector-vector response (V-V) is
\begin{align}
\left( \begin{matrix}
		j^\mu \\
		q^\mu \\
\end{matrix} \right)
= \tau 
\left(\Delta^{\mu\nu}+ \tau B_{(G)}^{\mu\nu}\right)
\left( \begin{matrix}
	 d^{-1}\mathcal{C}_{1} \\
	d^{-1}\mathcal{C}_{2} \\
\end{matrix} \right)
\mathcal D_\nu \beta .  \label{V-V}
\end{align}
\indent (iii) The tensor-tensor response (T-T) is
\begin{align}
{
\Pi^{\mu\nu}
= \tau \frac{2}{(d+2)d}
\left(\Delta^{\rho\mu}\Delta^{\nu\sigma}
+2\tau \Delta^{\rho(\mu} B_{(G)}^{\nu)\sigma} \right)\mathcal{C}_{2}\psi_{\sigma \rho}.} \label{T-T}
\end{align}
\indent (iv) The scalar-vector response (S-V) is
\begin{align}
&\left( \begin{matrix}
		\delta n \\
		\delta \epsilon \\
        \Pi
\end{matrix} \right)
=-
\tau^2 E_{(G)}^{\mu} 
\left( \begin{matrix}
	\mathcal{C}_{1}  \\
	   \mathcal{C}_{2}\\
    d^{-1} \mathcal{C}_{2} \\
\end{matrix} \right)\mathcal{D}_{\mu} \beta.\label{S-V} 
\end{align}
\indent (v) The vector-scalar response (V-S) is
\begin{align}
\left( \begin{matrix}
		j^\mu \\
		q^\mu \\
\end{matrix} \right)
=- \tau^2 
E_{(G)}^{\mu}
\left( \begin{matrix}
	 d^{-1}\mathcal{C}_{1} & d^{-2}\mathcal{C}_{1} \\
	 d^{-1}\mathcal{C}_{2} & d^{-2}\mathcal{C}_{2} \\
\end{matrix} \right)
\left( \begin{matrix}
		\dot \beta \\
        \psi \\
\end{matrix} \right). \label{V-S}
\end{align}
\indent (vi) The vector-tensor response (V-T) is
\begin{align}
\left( \begin{matrix}
		j^\mu \\
		q^\mu \\
\end{matrix} \right)
= -\tau^2 \frac{2}{d}
\left(1-\frac{1}{d+2}\right)E^{(G)}_{\rho}
\left( \begin{matrix}
	\mathcal{C}_{1}  \\
	\mathcal{C}_{2} \\
\end{matrix} \right)
\psi^{\rho\mu}. \label{V-T}
\end{align}
Here ``scalar-scalar response'' means that a scalar dissipative quantity responds to a scalar thermodynamic force, and the other response types are named analogously.

It follows from Eqs.~(\ref{S-S}--\ref{V-T}) that at first order in \(\tau\), each dissipative flux couples only to thermodynamic forces of the same tensorial character; the transport coefficients are therefore scalar. At second order, however, scalar dissipative terms can respond to vector forces, indicating that the gravitational field renders the transport coefficients tensor-valued. The standard Onsager-type constitutive relations \cite{Groot1968RelativisticOR, Hao_2024} are thus only valid at first order.

Similarly, Eq.~(\ref{V-V}) shows that at first order the fluxes are aligned with the thermodynamic force (longitudinal transport). At second order, transverse transport appears due to the antisymmetry of \(B^{(G)}_{\mu\nu}\), which produces flux components orthogonal to the driving force. The gravitoelectric field \(E^{(G)}_{\mu}\) induces additional effects. According to Eq.~(\ref{S-V}), it couples to the temperature gradient \(\mathcal{D}_\mu\beta\). In curved spacetime, global equilibrium requires the temperature to vary with the gravitational potential (the Tolman-Ehrenfest effect), therefore, when the gravitational potential changes, the equilibrium temperature, particle number density, and energy density change accordingly. During the non-equilibrium process in which the particle number density and energy density evolve from the equilibrium state prior to the change in gravitational potential to the new equilibrium state afterward, this coupling drives particle and energy redistribution along the gravitoelectric field.

Finally, unlike the massive particle system~\cite{Cui_2025_2}, for a photon gas, the ratio of energy density to particle number density (excluding zeroth-order terms) depends only on temperature and dimension. Likewise, in the fluid rest frame, the energy flux is parallel to the particle flux
\begin{equation}
    \frac{\epsilon^{(1)}}{n^{(1)}}=\frac{q^{\mu(1)}}{j^{\mu(1)}}=\frac{\epsilon^{(2)}_{\text{linear}}}{n^{(2)}_{\text{linear}}}=\frac{q^{\mu(2)}_{\text{linear}}}{j^{\mu(2)}_{\text{linear}}}.
\end{equation}
This reflects that the energy and particle transport are not independent.

We now derive the macroscopic transport laws from the linear response relations. In the non-relativistic regime, Fick's law and Fourier's law take the familiar forms
\begin{equation}
    \vec{j} = -D \vec{\nabla} n, \qquad \vec{q} = -\kappa \vec{\nabla} T.
\end{equation}
These formulas state that the particle flux and the energy flux are proportional to the gradients of density and temperature. However, the standard form of Fourier's law fails in the equilibrium limit of a curved spacetime, since it would require a uniform temperature whenever the energy flux vanishes, in contradiction with the Tolman-Ehrenfest condition, which allows a non-uniform temperature in curved spacetime equilibrium. This necessitates a covariant generalization.

To derive such generalizations from the linear response results, we note that in the near-equilibrium regime, the higher-order gradient contributions are much smaller than the first-order ones. It therefore suffices to retain only the lowest-order terms, which correspond to the familiar forms of Fick's and Fourier's laws. In the RTA framework, these correspond to the \(\mathcal{O}(\tau)\) parts of the vector-vector response, while the \(\mathcal{O}(\tau^2)\) corrections contribute at higher order in the gradient expansion.

Starting from the first-order vector-vector response in Eq.~(\ref{V-V}), since \(n^{(0)}\) has the same functional dependence on \(T\) as the equilibrium energy density in Eq.~(\ref{functions of temperature}), we have \[
\frac{\nabla_\nu n^{(0)}}{\nabla_\nu T} = \frac{d n^{(0)}}{T}=\frac{\mathcal{C}_1}{T^2}.
\]
We obtain
\begin{equation}
    \begin{split}
        j^{\mu(1)} &= -D_{R}\left(\nabla_{\nu}n^{(0)}+d\,n^{(0)}U^{\sigma}\nabla_{\sigma} U_{\nu}\right)\Delta^{\mu\nu},\\
        q^{\mu(1)} &= -\kappa_{R}\left(\nabla_{\nu}T+T\,U^{\sigma}\nabla_{\sigma} U_{\nu}\right)\Delta^{\mu\nu},
    \end{split}
    \label{Fourier law}
\end{equation}
with
\begin{equation}
    D_{R}\equiv\frac{\tau}{d},\qquad
    \kappa_{R}\equiv\frac{\tau}{d}\frac{\mathcal{C}_2}{T^2}.
\end{equation}
These are the relativistic generalizations of Fick's and Fourier's laws for a photon gas. Compared to their non-relativistic counterparts, they contain additional acceleration terms \(U^\sigma\nabla_\sigma U_\nu\) arising from spacetime curvature, which ensure that the fluxes vanish in equilibrium when the Tolman-Ehrenfest condition holds. 

This formulation provides the constitutive relations needed to close the conservation equations for the photon gas. In the next section, we substitute the generalized Fourier law into the energy-momentum conservation equation to obtain the radiation diffusion equation, which we then apply to spherically symmetric accretion disks around Schwarzschild black holes.

\section{Radiation Diffusion Equation at the Spherical Accretion Disk}
In high-energy astrophysical environments such as accretion disks and massive stars, radiation is the dominant mechanism of energy transport and cooling. The emitted photons carry information about the underlying physical processes and constitute the primary observational signal. Modeling this radiative transport in strong gravitational fields therefore requires a fully relativistic, observer-independent formulation.

In order to obtain the evolution of the energy density distribution of the photon system in high-energy astrophysical environments, we can obtain the 1PDF via the RTA, then substitute it into Eq.(\ref{int}) to compute the energy-momentum tensor of the system, and finally obtain the energy density. However, this procedure is too cumbersome and complicated. In the preceding sections, we have already established the linear response theory for the photon system in such high-energy astrophysical environments, which can describe the transport behavior of energy in non-equilibrium systems such as stellar interiors or black hole accretion disks. By means of the linear response theory for the photon system and the conservation equation of the energy-momentum tensor, one can directly construct the differential equation governing the evolution of the energy density.

The classical radiation diffusion equation follows from energy conservation together with Fourier's law. In general relativity, the corresponding conservation law is the covariant conservation of the total energy-momentum tensor given by
\begin{equation}
    U_{\nu}\nabla_{\mu}(T^{\mu \nu}+T_r^{\mu \nu})=0 ,\label{energy conservation equation}
\end{equation}
where \(T^{\mu\nu}\) is the photon energy-momentum tensor and \(T_r^{\mu\nu}\) is that of the background heat reservoir. As a simple approximation, we consider a steady-state accretion disk in a stationary spacetime. The background fluid evolves on a timescale much longer than that of photon transport, so it can be treated as static during photon propagation. Equation (\ref{energy conservation equation}) therefore reduces to the photon energy conservation equation,
\begin{equation}
    U_{\nu}\nabla_{\mu}T^{\mu \nu}=0. \label{energy conservation equation of the photon}
\end{equation}

Substituting the non-equilibrium decomposition in Eq.~(\ref{non-equilibrium state2}) into Eq.~(\ref{energy conservation equation of the photon}) and keeping terms up to first order in the relaxation time, with \(\delta\epsilon=\epsilon^{(1)}\), yields
\begin{equation}
    U^{\mu}\nabla_{\mu}(\epsilon^{(0)}+\epsilon^{(1)})+(P+\Pi^{(1)}+\epsilon^{(0)}+\epsilon^{(1)})\nabla_{\mu}U^{\mu}+\nabla_{\mu}q^{\mu(1)}+q^{\nu(1)}U^{\mu}\nabla_{\mu}U_{\nu}=0.
\end{equation}
For simplicity, we focus on photon transport driven by temperature gradients. We assume the fluid four-velocity is approximately aligned with a timelike Killing vector field, so that the expansion vanishes, \(\nabla_\mu U^\mu = 0\). For the static fluid configurations considered here, this is reasonable. Since the system is only slightly out of equilibrium, the first-order correction to the energy density \(\epsilon^{(1)}\) is much smaller than its zeroth-order value \(\epsilon^{(0)}\) and is neglected. Note that this does not affect the dissipative fluxes \(q^{\mu(1)}\) and \(\Pi^{(1)}\), which have no zeroth-order counterparts and must be retained to describe transport.

For later convenience, it is useful to rewrite the generalized Fourier law in Eq.~(\ref{Fourier law}) in terms of the equilibrium energy density \(\epsilon^{(0)}\). Since \(\epsilon^{(0)}\) has the same functional dependence on \(T\) as the equilibrium energy density in Eq.~(\ref{functions of temperature}), we have
\[
\frac{\nabla_\nu \epsilon^{(0)}}{\nabla_\nu T} = \frac{(d+1)\epsilon^{(0)}}{T}=\frac{\mathcal{C}_2}{T^2}.
\]
Using this relation and the definition of \(D_R\) and \(\kappa_R\), Eq.~(\ref{Fourier law}) becomes
\begin{equation}
    q^{\mu(1)}=-D_{R}\left[\nabla_{\nu}\epsilon^{(0)}+(d+1)\epsilon^{(0)}U^{\sigma}\nabla_{\sigma} U_{\nu}\right]\Delta^{\mu\nu}.
    \label{Fourier's law2}
\end{equation}
Inserting this into the photon energy conservation equation then yields the radiation diffusion equation,
\begin{equation}
     \begin{split}
          &U^{\mu}\nabla_{\mu}\epsilon^{(0)}-D_R \nabla_{\mu}\left[(\nabla_{\nu}\epsilon^{(0)}+(d+1)\epsilon^{(0)}A_{\nu})\Delta^{\mu \nu}\right]-D_R[\nabla_{\mu}\epsilon^{(0)}+(d+1)\epsilon^{(0)}A_{\mu}]A^{\mu}=0,
     \end{split} \label{radiation diffusion equation general}
\end{equation}
where \(A^{\nu}=U^{\mu}\nabla_{\mu}U^{\nu}\) is the covariant acceleration.

We now specialize to a spherically symmetric accretion disk around a Schwarzschild black hole, setting the spatial dimension to \(d=3\). The black hole mass is denoted by \(M\) and the gravitational constant by \(G\). The fluid four-velocity is taken to be
\begin{equation}
   \begin{split}
       U^{\mu}=\left(\frac{1}{\sqrt{1-\dfrac{2GM}{r}}},0,0,0\right).
   \end{split}
\end{equation}
For the Schwarzschild metric, the acceleration is purely radial, \(A^r = -GM/r^2\), and the projection operator gives \(\Delta^{rr} = 1 - 2GM/r\). Substituting these into Eq.~(\ref{radiation diffusion equation general}) and imposing spherical symmetry yields
\begin{equation}
 \begin{split}
     &\frac{3}{\sqrt{1-\dfrac{2GM}{r}}}\frac{\partial \epsilon^{(0)}}{\partial t}-\tau\frac{r-2GM}{r}\frac{\partial^{2} \epsilon^{(0)}}{\partial r^2}-\tau\frac{6GM+4r}{2r^2}\frac{\partial \epsilon^{(0)}}{\partial r}-\tau\frac{4G^2M^2}{r^3(r-2GM)}\epsilon^{(0)}=0.
     \label{radiation diffusion equation}
 \end{split}
\end{equation}
 \(\epsilon^{(0)}\) describes the distribution of the zeroth‑order energy density of the spherical accretion disk, which is a function of the time coordinate \(t\) and the spatial coordinate \(r\), i.e., \(\epsilon^{(0)}=\epsilon^{(0)}(t,r)\). The only parameter in this equation that depends on the composition of the disk is the relaxation time \(\tau\), which encodes the microphysics of the photon-matter interactions. The equation is therefore valid whenever the relaxation time approximation provides an adequate description of the system.

\begin{figure}
    \centering
    \includegraphics[width=0.5\linewidth]{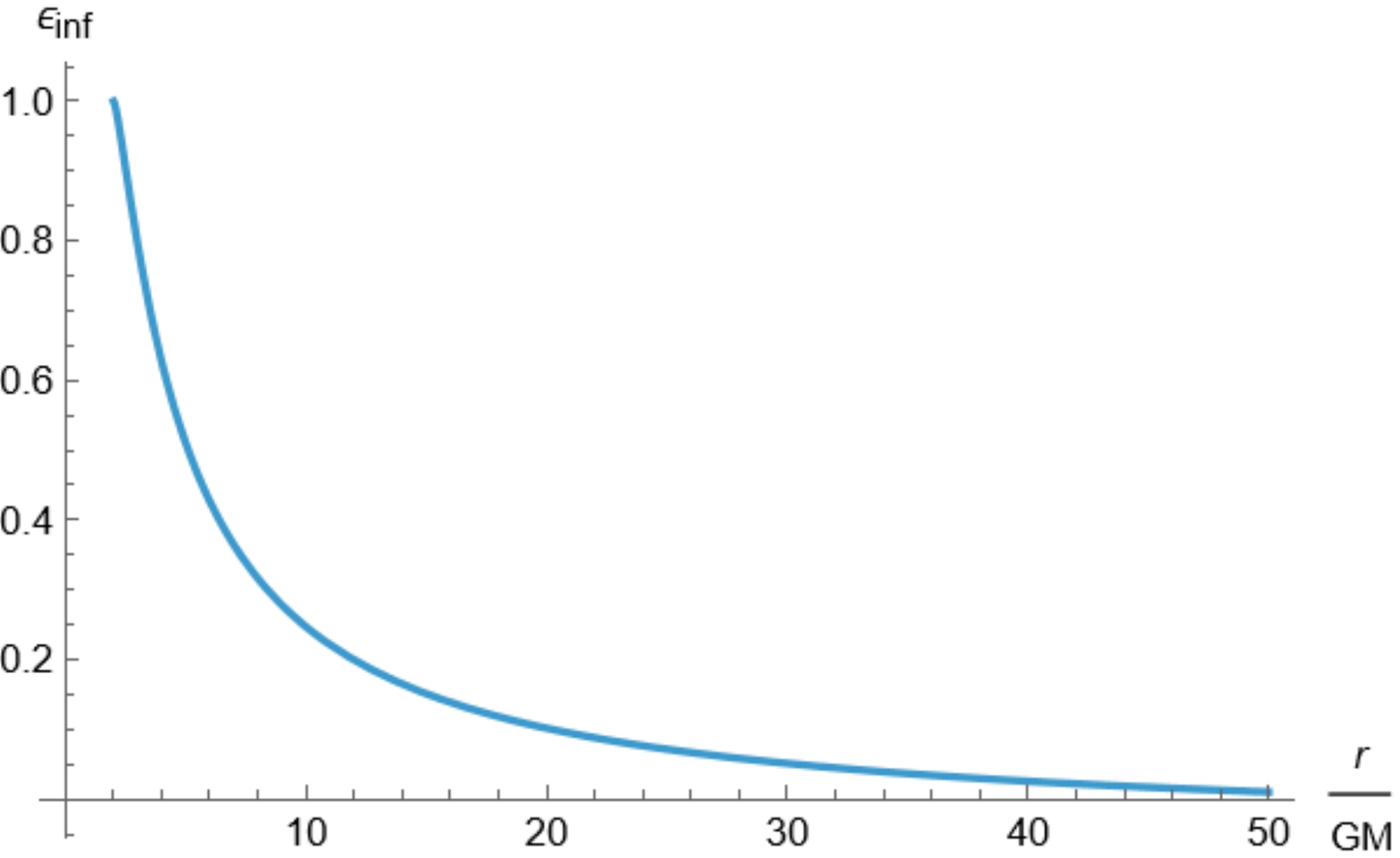}
    \caption{Steady-state solution of the radiation diffusion equation}
    \label{fig:placeholder}
\end{figure}

For a Schwarzschild black hole of mass \(M\), consider the vicinity of the event horizon where 
\begin{equation}
1-\frac{2GM}{r} \ll 1. 
\end{equation}
The global equilibrium condition (\ref{detailed balance condition}) implies that the equilibrium temperature is inversely proportional to the norm of the timelike Killing vector field. Near the horizon, this norm vanishes, so both the equilibrium temperature and the equilibrium energy density diverge. This divergence is a gravitational redshift effect. To eliminate it in numerical simulations, we define a redshifted energy density \(\epsilon_{\mathrm{inf}}^{(0)}\) as the energy density measured by an observer at infinity,
\begin{equation}
    \epsilon^{(0)}_{\mathrm{inf}} = \epsilon^{(0)}\left(1-\frac{2GM}{r}\right)^2.
    \label{gravitational redshift factor}
\end{equation}
As a consistency check, note that if \(\epsilon^{(0)}_{\text{inf}}\) is constant, Eq.~(\ref{Fourier's law2}) gives \(q^{\mu(1)}=0\), confirming that the energy flux vanishes in equilibrium.
Substituting Eq.~(\ref{gravitational redshift factor}) into Eq.~(\ref{radiation diffusion equation}) yields
\begin{equation}
 \begin{split}
       \frac{3}{\sqrt{1-\dfrac{2GM}{r}}}\frac{\partial \epsilon^{(0)}_{\mathrm{inf}}}{\partial t}-\tau\left(1-\frac{2GM}{r}\right)\frac{\partial^2\epsilon^{(0)}_{\mathrm{inf}}}{\partial r^2}-\tau\frac{2r-5GM}{r^2}\frac{\partial \epsilon^{(0)}_{\mathrm{inf}}}{\partial r}=0.
       \label{radiation diffusion equation2}
 \end{split}
\end{equation}
In the absence of sources, a constant is the simplest steady-state solution, since any spatial gradient would drive a diffusive flux that continues until the gradient vanishes. However, when a stable energy source is present near the horizon, for instance gravitational energy released by accretion, the steady-state profile is determined by the boundary conditions \(\epsilon^{(0)}_{\mathrm{inf}} = 1\) at the horizon and \(\epsilon^{(0)}_{\mathrm{inf}} = 0\) at the outer edge of the disk. The resulting profile is shown in Fig.~(\ref{fig:placeholder}), where the horizontal axis is the radius in units of \(GM\), and the vertical axis is the redshifted energy density \(\epsilon^{(0)}_{\mathrm{inf}}\).

Physically, the profile reflects a balance between the inward gravitational redshift and the outward radiative diffusion. The energy density is highest near the horizon, where the source is located, and decreases monotonically toward the outer boundary as photons diffuse outward. This behavior is characteristic of radiation-dominated accretion flows. The gravitational field confines the radiation, while diffusion transports it outward.

\section{Discussion, Summary and Outlook} 

Here we compare our expansion in powers of the relaxation time \(\tau\) with the 14-moment formalism of Udey and Israel~\cite{Udey1982}. Both approaches start from the general-relativistic Boltzmann equation, adopt the relaxation-time approximation for the collision term, and rely on the local-equilibrium assumption. They differ in how the distribution function is expanded and in how the dissipative quantities are treated.

\begin{itemize}
\item[(i)] \emph{Expansion and dissipative variables.} 14-moment approach expands the distribution function in momentum moments, truncating the hierarchy at the 14-moment level, and treats the dissipative quantities as independent dynamical variables obeying their own equations of motion. We instead expand the distribution function iteratively in powers of \(\tau\). The dissipative quantities are then given by constitutive relations in terms of the hydrodynamic variables \((T,U^\mu)\), and their evolution follows entirely from the fluid conservation equations.

\item[(ii)] \emph{Hydrodynamic frame.} Their formalism adopts the frame of the material medium. The four-velocity is that of the matter, and the equilibrium state is fitted to the actual state by matching the total energy density and the baryon density to their equilibrium values, so that the radiative dissipative fluxes are defined relative to the medium. Our scheme imposes no such matching conditions. $U^{\mu}$ and \(T\) remain those of the photon gas and a particular hydrodynamic frame can be selected afterward.

\item[(iii)] \emph{Tensor-type transport coefficients.} At second order in \(\tau\), our method yields terms with no counterpart in the standard 14-moment truncation. The transport coefficients become tensorial, the gravitomagnetic field $B^{(G)}_{\mu\nu}$ produces fluxes orthogonal to the driving force, and the gravitoelectric field $E^{(G)}_{\mu}$ couples vector and scalar fluxes (the S--V and V--S responses). This difference stems from the closure. The 14-moment approach discards rank-three coefficients, thereby eliminating the three-index momentum structures that survive in our \(f^{(2)}\). While the 14-moment approach also yields transverse couplings to the fluid vorticity, \(B^{(G)}_{\mu\nu}\) is physically equivalent to this vorticity in the medium frame. The key distinction lies in where the coupling acts. The 14-moment approach includes it in the equation of motion of the dissipative quantities, whereas our scheme incorporates it into the constitutive relation at second order.
\end{itemize}

The resulting transport coefficients are explicitly compared in Appendix~\ref{appendixA}. For \(d=3\) and to first order in \(\tau\), our thermal conductivity agrees with that of the 14-moment approach, and our shear coefficient agrees once the factor of 2 in the normalization of the shear stress is taken into account. The bulk viscous coefficients, however, refer to different objects. In our case, the bulk viscous coefficient describes the response of the dynamic pressure to the expansion scalar of the photon gas alone, while in the 14-moment approach it describes the response of the combined radiation–matter system.

We now summarize the main results in this work. We have developed a covariant linear-response theory for a relativistic photon gas in curved spacetime. Treating the ambient plasma as a thermal reservoir, we computed the photon transport coefficients up to second order in the relaxation time using the relativistic Boltzmann equation within the relaxation time approximation. At first order in the relaxation time, the transport coefficients are scalar and describe standard diffusion and heat conduction. At second order, the gravitational field, encoded in the gravitoelectric and gravitomagnetic fields \(E^{(G)}_{\mu}\) and \(B^{(G)}_{\mu\nu}\), induces two novel effects. Importantly, these photon transport coefficients are not a massless limit of the massive-particle results, as the phase-space structure and the absence of particle number conservation are fundamentally different for photons.

Using the first-order transport coefficients, we derived the relativistic generalizations of Fick's and Fourier's laws, which contain additional acceleration terms arising from spacetime curvature. Substituting the generalized Fourier law into the energy-momentum conservation equation yielded a covariant radiation diffusion equation. As an application, we specialized to a spherically symmetric accretion disk around a Schwarzschild black hole and obtained the explicit form of the diffusion equation. Numerical solution under appropriate boundary conditions gave the steady-state profile of the redshifted energy density, which we presented in Fig.~(\ref{fig:placeholder}).

While the relaxation time approximation provides a tractable framework for computing transport coefficients, it does not capture the detailed frequency dependence of photon-matter interactions. A complete description would require solving the full Boltzmann collision integral, which remains a direction for future work. Further extensions include generalizing the diffusion equation to rotating black hole spacetimes, where the gravitomagnetic field may play a more prominent role, and exploring possible observational signatures of the transverse transport effects in X-ray spectra or polarization of accreting systems.

\section*{Acknowledgment}
This work is supported by the National Natural Science Foundation of China (Grant Nos.12275216, 12247103).

\begingroup\raggedright\endgroup

\appendix 
\section{First Order in Relaxation Time} \label{appendixA}
Substituting Eq.(\ref{first-order}) into the integral expressions for the particle current vector and the energy-momentum tensor (\ref{int}), and using Eq.(\ref{momentum}) and Eq.(\ref{volume element}), the particle current vector of the first-order can be derived
\begin{align}
    N^{\mu(1)}&=\int{\varpi p^{\mu}f^{(1)}}\notag\\
            &=-\tau \int \varpi\frac{1}{\varepsilon}\nabla_{(\nu}\mathcal{B}_{\sigma)}p^{\mu}p^{\nu}p^{\sigma}\frac{\partial f^{(0)}}{\partial \xi}\notag\\
            &=-\tau \int \mathrm{d}p^{0} \mathrm{d}\Omega_{d-1} (p^0)^{d}\frac{\partial f^{(0)}}{\partial \xi}\nabla_{(\nu}\mathcal{B}_{\sigma)}(U^{\mu}U^{\nu}U^{\sigma}+3n^in^jU^{(\mu}{\Delta_{i}}^{\nu}{\Delta_j}^{\sigma)})\notag\\
            &=\tau\frac{\mathfrak{g}_{0}}{h^d}\Omega_{d-1}\Gamma(d+1)\zeta(d)T^{d+1}\left(\dot{\beta }U^{\mu}+\frac{1}{d}U^{\mu}\psi+\frac{1}{d}\Delta^{\mu\sigma}\mathcal{D}_{\sigma}\beta\right),
\end{align}
 and the energy-momentum tensor of the first-order can be derived
\begin{align}
      T^{\mu\nu(1)}&=\int{\varpi p^{\mu}p^{\nu}f^{(1)}}\notag\\ 
                &=-\tau\int \varpi\frac{1}{\varepsilon}\nabla_{(\rho}\mathcal{B}_{\sigma)}p^{\mu}p^{\nu}p^{\rho}p^{\sigma}\frac{\partial f^{(0)}}{\partial \xi}\notag\\
                &=-\tau \int  \mathrm{d}p^{0} \mathrm{d}\Omega_{d-1} (p^0)^{d+1}\frac{\partial f^{(0)}}{\partial \xi}\nabla_{(\nu}\mathcal{B}_{\sigma)}\left(U^{\mu}U^{\nu}U^{\rho}U^{\sigma}+\frac{6}{d}U^{(\mu}U^{\nu}\Delta^{\rho \sigma)}+\frac{3}{d(d+2)}\Delta^{(\mu\nu}\Delta^{\rho\sigma)}\right)\notag\\
                &=\tau \frac{\mathfrak{g}_{0}}{h^d}\Omega_{d-1}\Gamma(d+2)\zeta(d+1)T^{d+2}\bigg(\dot{\beta}U^{\mu}U^{\nu}+\frac{1}{d}\dot{\beta}\Delta^{\mu\nu}+\frac{1}{d}\psi U^{\mu}U^{\nu}+\frac{2}{d}U^{(\mu}\Delta^{\nu)\rho}\mathcal{D}_{\rho}\beta\notag\\
                &\quad+\frac{1}{d^2}\Delta^{\mu\nu}\psi+\frac{2}{d(d+2)}\Delta^{\mu(\rho}\Delta^{\sigma)\nu}\psi_{\rho\sigma}\bigg)
\end{align}
Comparing these results with the macroscopic expressions for the particle current and energy-momentum tensor in non-equilibrium states (\ref{non-equilibrium state1} and \ref{non-equilibrium state2}) yields the linear relations between the variations of macroscopic thermodynamic quantities of the photon system and the thermodynamic forces. The coefficients in these linear relations are precisely the transport coefficients of the system. These transport coefficients can be obtained to first order in the relaxation time,
\begin{align}
        n^{(1)}&=-U_{\mu}N^{\mu(1)}\notag\\
               &=\tau\frac{\mathfrak{g}_{0}}{h^d}\Omega_{d-1}\Gamma(d+1)\zeta(d)T^{d+1}\left(\dot{\beta}+\frac{1}{d}\psi\right),\notag\\
        \epsilon^{(1)}&=U_{\mu}U_{\nu}T^{\mu\nu(1)}\notag\\
               &=\tau\frac{\mathfrak{g}_{0}}{h^d}\Omega_{d-1}\Gamma(d+2)\zeta(d+1)T^{d+2}\left(\dot{\beta}+\frac{1}{d}\psi\right),\notag\\
        \Pi^{(1)}&=\frac{1}{d}\Delta_{\mu\nu}T^{\mu\nu(1)}\notag\\
               &=\tau\frac{\mathfrak{g}_{0}}{h^d}\Omega_{d-1}\Gamma(d+2)\zeta(d+1)T^{d+2}\left(\frac{1}{d}\dot{\beta}+\frac{1}{d^2}\psi\right),\notag\\
        j^{\mu(1)}&={h^{\mu}}_{\nu}N^{\nu(1)}\label{transport coefficients1}\\
               &=\tau\frac{\mathfrak{g}_{0}}{h^d}\Omega_{d-1}\Gamma(d+1)\zeta(d)T^{d+1}\left(\frac{1}{d}\Delta^{\mu\nu}\mathcal{D}_{\nu}\beta\right),\notag\\
        q^{\mu(1)}&=-U_{\nu}{h^{\mu}}_{\sigma}T^{\nu\sigma(1)}\notag\\
               &=\tau\frac{\mathfrak{g}_{0}}{h^d}\Omega_{d-1}\Gamma(d+2)\zeta(d+1)T^{d+2}\left(\frac{1}{d}\Delta^{\mu\nu}\mathcal{D}_{\nu}\beta\right),\notag\\
        \Pi^{\mu\nu(1)}&={\Delta^{\mu}}_{\rho}{\Delta^{\nu}}_{\sigma}T^{\rho\sigma(1)}-\frac{1}{d}\Delta^{\mu\nu}\Delta_{\rho\sigma}T^{\rho\sigma(1)}\notag\\
               &=\tau\frac{\mathfrak{g}_{0}}{h^d}\Omega_{d-1}\Gamma(d+2)\zeta(d+1)T^{d+2}\left(\frac{2}{d(d+2)}\Delta^{\rho\mu}\Delta^{\nu\sigma}\psi_{\rho\sigma}\right).\notag
\end{align}

Furthermore, to first order in the relaxation time, using Eq. (\ref{T-T}) and Eq. (\ref{S-S}), we obtain the shear coefficient $\zeta_S$ and bulk coefficient \(\zeta_V\) of the photon gas. For a constant temperature profile (\(\dot{\beta}=0\)), the dissipative fluxes reduce to
\begin{equation}
    \begin{split}
        &\Pi^{\mu\nu} \equiv -\zeta_S\left({\Delta_{(\mu}}^{\rho}{\Delta_{\nu)}}^{\sigma}\nabla_{\rho}U_{\sigma}-\frac{1}{d}\nabla_\rho U^\rho\Delta^{\mu\nu}\right) ,\quad\Pi\equiv-\zeta_V\nabla_\rho U^\rho,\\
        &\zeta_S=\frac{2\tau\mathfrak{g}_{0}\Omega_{d-1}\Gamma(d+2)\zeta(d+1)T^{d+1}}{h^dd(d+2)},\\
        &\zeta_V=\frac{\tau\mathfrak{g}_{0}\Omega_{d-1}\Gamma(d+2)\zeta(d+1)T^{d+1}}{h^dd^2}.
        \label{shear  coefficient}
    \end{split}
\end{equation}

For comparison with the 14-moment results of Udey and Israel~\cite{Udey1982}, who work in four-dimensional spacetime, we specialize our results to \(d=3\). Their results read
\begin{equation}
    \kappa^{\rm UI}=\frac{4}{3}aT^3\tau,\qquad
    \zeta_S^{\rm UI}=\frac{4}{15}aT^4\tau,\qquad
    \zeta_V^{\rm UI}=12 aT^4\Omega^2\tau,
    \label{UdeyIsrael54}
\end{equation}
with \(\Omega\equiv\frac{1}{3}-(\partial P_{\rm tot}/\partial\rho_{\rm tot})_n\) and \(a=8\pi^5/(15h^3)\). Our results in Eqs.~(\ref{Fourier law}) and~(\ref{shear coefficient}) reduce, for \(d=3\), to
\begin{equation}
    \kappa_R=\frac{4}{3}aT^3\tau,\qquad
    \zeta_S=\frac{64\pi^5}{225h^3}\tau T^4=\frac{8}{15}aT^4\tau,\qquad
    \zeta_V=\frac{32\pi^5}{135h^3}\tau T^4=\frac{4}{9}aT^4\tau.
    \label{ours d=3}
\end{equation}
The thermal conductivity agrees directly, \(\kappa_R=\kappa^{\rm UI}\). For the shear coefficient, the two papers use different normalization. They write the shear stress as \(-2\zeta_S\sigma^{\mu\nu}\) (their Eq.~(46)), whereas we use \(\Pi^{\mu\nu}=-\zeta_S\sigma^{\mu\nu}\). Taking this factor of 2 into account, our result satisfies \(\zeta_S=2\zeta_S^{\rm UI}=\frac{8}{15}aT^4\tau\), so the two agree once the normalizations are matched. The bulk viscous coefficient, by contrast, is defined differently. 

The quantity \(\zeta_V^{\rm UI}\) is the bulk viscosity of the combined radiation--matter system, obtained after subtracting the reversible thermodynamic pressure (their Eqs.~(27)--(33)), and therefore depends on the equation of state through \(\Omega\), whereas our \(\zeta_V\) is the bulk stress of the photon gas alone, arising from the response of the dynamic pressure to the scalar thermodynamic forces. In the limit \(\frac{\partial P_{\mathrm{tot}}}{\partial\rho_{\mathrm{tot}}}\to 0\) (i.e., \(\Omega = 1/3\)), the matter-radiation thermal coupling vanishes, isolating the intrinsic kinetic contribution. The Udey-Israel coefficient then evaluates to \(\zeta_V^{\mathrm{UI}} = \frac{4}{3} aT^4\tau\), which is 3 times our \(\zeta_V = \frac{4}{9} aT^4\tau\). This factor of 3 arises entirely from the normalization convention of the bulk stress (their \(\pi = -\frac{1}{3}\zeta_V^{\mathrm{UI}}\theta\) versus our \(\Pi = -\zeta_V\theta\)). Once we compare the effective stress coefficients, both approaches yield exactly \(\frac{4}{9}aT^{4}\tau\).

\section{Second Order in Relaxation Time}\label{appendixB}
In order to obtain the linear response term of the second-order relaxation time, we need to examine them term by term. The gravitoelectromagnetic field can be obtained from the decomposition of the following important decomposition:
\begin{equation}
    \begin{split}
        \nabla_{\mu} U_{\nu}  = &~ \Delta^{\rho}_{\enspace \mu}
  \Delta^{\sigma}_{\enspace \nu} \nabla_{(\rho} U_{\sigma)} +
  \Delta^{\rho}_{\enspace \mu} \Delta^{\sigma}_{\enspace \nu} \nabla_{[\rho}
  U_{\sigma]} - U_{\mu} U^{\rho} \nabla_{\rho} U_{\nu} \\
   = & - \frac{1}{\beta} \psi_{\mu \nu} - \frac{1}{d}  \frac{1}{\beta} \psi
  \Delta_{\mu \nu} + B_{\mu \nu}^{(G)} +  U_{\mu} E^{(G)}_{\nu}
  \label{U}
    \end{split}
\end{equation}
By using Eqs.(\ref{decomposition}) and (\ref{U}), we can decompose $\nabla_{\sigma} (\nabla_{(\rho}\mathcal{B}_{\lambda)})$
in Eq.(\ref{2t}) into linear response terms, derivative terms, and product terms of thermodynamic forces,
\begin{align}
        \nabla_{\sigma} (\nabla_{(\rho}\mathcal{B}_{\lambda)}) = & - (\nabla_{\sigma}
  \dot{\beta}) U_{\rho} U_{\lambda}\notag\\
  &-  2\dot{\beta}  \left( -
  \frac{1}{\beta} \psi_{\sigma (\rho} U_{\lambda)} - \frac{1}{d}
  \frac{1}{\beta} \psi \Delta_{\sigma (\rho} U_{\lambda)} + B_{\sigma
  (\rho}^{(G)} U_{\lambda)} +  U_{\sigma} E^{(G)}_{(\rho}
  U_{\lambda)} \right)\notag\\
    & + \nabla_{\sigma} (\mathcal{D}_{(\rho}\beta )U_{\lambda)} -
  \frac{1}{\beta} \psi_{\sigma (\lambda} \mathcal{D}_{\rho)} \beta -
  \frac{1}{d}  \frac{1}{\beta} \psi \Delta_{\sigma (\lambda}
  \mathcal{D}_{\rho)} \beta \\
  &+ B_{\sigma (\lambda}^{(G)} \mathcal{D}_{\rho)}
  \beta +  U_{\sigma} E^{(G)}_{(\lambda} \mathcal{D}_{\rho)}
  \beta - \nabla_{\sigma} \psi_{\rho \lambda} - \frac{1}{d} (\nabla_{\sigma}
  \psi) \Delta_{\rho \lambda} \notag\\
  &- \frac{2}{d} \psi  \left( -
  \frac{1}{\beta} \psi_{\sigma (\rho} U_{\lambda)}- \frac{1}{d}
  \frac{1}{\beta} \psi \Delta_{\sigma (\rho} U_{\lambda)} + B_{\sigma
  (\rho}^{(G)} U_{\lambda)} +  U_{\sigma} E^{(G)}_{(\rho}
  U_{\lambda)} \right),\notag
\end{align}
where $\nabla_{\sigma} \mathcal{D}_{\alpha} \beta$, and
$\nabla_{\sigma} \psi_{\rho \lambda}$ can be further decomposed as:
\begin{equation}
    \begin{split}
        \nabla_{\sigma} \mathcal{D}_{\alpha} \beta  = &~ \Delta^{\gamma}_{\enspace
  \sigma} \Delta^{\beta}_{\enspace \alpha} \nabla_{\gamma} (\mathcal{D}_{\beta}
  \beta) - U_{\sigma} \Delta^{\beta}_{\enspace \alpha} U^{\gamma}
  \nabla_{\gamma}( \mathcal{D}_{\beta} \beta) \\
  &+U_{\alpha}
  \Delta^{\beta \gamma} \left( B^{(G)}_{\sigma \gamma} + 
  U_{\sigma} E^{(G)}_{\gamma} \right) \mathcal{D}_{\beta} \beta\ +  U_{\alpha} (\mathcal{D}_{\beta} \beta )\Delta^{\beta
  \gamma}\Delta^{\delta}_{\enspace \sigma} \nabla_{(\gamma} U_{\delta)},\\
    \end{split}
\end{equation}
\begin{equation}
    \begin{split}
         \nabla_{\sigma} \psi_{\rho \lambda}  = &~ \Delta^{\alpha}_{\enspace \sigma}
  \Delta^{\beta}_{\enspace \rho} \Delta^{\gamma}_{\enspace \lambda}
  \nabla_{\alpha} \psi_{\beta \gamma} -  U_{\sigma} U^{\alpha}
  \Delta^{\beta}_{\enspace \rho} \Delta^{\gamma}_{\enspace \lambda}
  \nabla_{\alpha} \psi_{\beta \gamma}\\
    & + 2\psi_{\beta \gamma} \Delta^{\eta \beta}
  \Delta^{\gamma}_{\enspace (\lambda} U_{\rho)} \left( - \frac{1}{\beta}
  \psi_{\sigma \eta} - \frac{1}{d}  \frac{1}{\beta} \psi \Delta_{\sigma \eta}
  + B_{\sigma \eta}^{(G)} +  U_{\sigma} E^{(G)}_{\eta} \right).
    \end{split}
\end{equation}
Then, by neglecting the nonlinear terms in the thermodynamic forces,the linear response term of the second-order can be obtained
\begin{align}
          f^{(2)}_{\text{linear}}=&-\left( \frac{\tau}{\varepsilon} \right)^{2}p^{\sigma}p^{\rho}p^{\lambda}\bigg[ - 2
  \dot{\beta}  \left( B_{\sigma \rho}^{(G)} U_{\lambda} + 
  U_{\sigma} E^{(G)}_{\rho} U_{\lambda} \right)-\frac{2}{d} \psi
   \left( B_{\sigma \rho}^{(G)} U_{\lambda} + 
  U_{\sigma} E^{(G)}_{\rho} U_{\lambda} \right) \notag\\
    & + B_{\sigma \lambda}^{(G)} \mathcal{D}_{\rho} \beta +U_{\sigma}E^{(G)}_{\rho}D_{\lambda}\beta+\Delta^{\beta \alpha} \left( B^{(G)}_{\sigma \alpha} + 
  U_{\sigma} E^{(G)}_{\alpha} \right) \mathcal{D}_{\beta} \beta \, U_{\rho}
  U_{\lambda} -E^{(G)}_{\sigma}\nabla_{\rho}\mathcal{B}_{\lambda}\label{linear part2}\\
    & - 2 \psi_{\beta \gamma} \Delta^{\eta \beta}
  \Delta^{\gamma}_{\enspace \lambda} U_{\rho} \left( B_{\sigma \eta}^{(G)} +
   U_{\sigma} E^{(G)}_{\eta} \right) \bigg].\notag
\end{align}
Similarly, using Eq.(\ref{int}), The particle current vector of the second-order can be derived:
\begin{equation}
    \begin{split}
        N^{\mu(2)}_{\text{linear}}&=\int\varpi p^{\mu}f^{(2)}_{\text{linear}}\\
        &=\tau^2 \frac{\mathfrak{g}_0}{h^d}\Omega_{d-1} \Gamma(d+1)\zeta(d)T^{d+1} \bigg[-\frac{1}{d}\dot{\beta}E^{\mu}_{(G)}+\frac{1}{d}B^{\mu\nu}_{(G)}\mathcal{D}_{\nu}\beta\\
        &\quad-\left(\frac{2}{d}-\frac{2}{d(d+2)}\right)\Delta^{\mu\rho}E^{\sigma}_{(G)}\psi_{\rho \sigma}-\frac{1}{d^2}\psi E^{\mu}_{(G)}-U^{\mu}E^{\rho}_{(G)}\mathcal{D}_{\rho}\beta \bigg],\\
    \end{split}
\end{equation}
and the energy-momentum tensor of the second-order can be derived:
\begin{equation}
    \begin{split}
        T^{\mu\nu(2)}_{\text{linear}}&=\int\varpi p^{\mu}p^{\nu}f^{(2)}_{\text{linear}}\\
        &=\tau^2 \frac{\mathfrak{g}_0}{h^d}\Omega_{d-1} \Gamma(d+2)\zeta(d+1)T^{d+2} \bigg[-\frac{2}{d}\dot{\beta}U^{(\mu}E^{\nu)}_{(G)}+\frac{2}{d}U^{(\mu}B^{\nu)\rho}_{(G)}\mathcal{D}_{\rho}\beta\\&\quad-\frac{1}{d}\Delta^{\mu\nu}E^{\rho}_{(G)}\mathcal{D}_{\rho}\beta-\frac{4}{d}\left(1-\frac{1}{d+2}\right)U^{(\mu}\Delta^{\nu)\rho}E^{\sigma}_{(G)}\psi_{\rho\sigma}-\frac{2}{d^2}U^{(\mu}E^{\nu)}_{(G)}\psi\\
        &\quad+\frac{4}{d(d+2)}B^{\rho(\mu}_{(G)}\Delta^{\nu)\sigma}\psi_{\rho\sigma}-U^{\mu}U^{\nu}E^{\rho}_{(G)}\mathcal{D}_{\rho}\beta\bigg].
    \end{split}
\end{equation}
The transport coefficients accurate to second order in the relaxation time,
\begin{align}
        n^{(2)}_{\text{linear}}&=-U_{\mu}N^{\mu(2)}_{\text{linear}}\notag\\
               &=\tau^2\frac{\mathfrak{g}_{0}}{h^d}\Omega_{d-1}\Gamma(d+1)\zeta(d)T^{d+1}(-E^{\rho}_{(G)}\mathcal{D}_{\rho}\beta),\notag\\
        \epsilon^{(2)}_{\text{linear}}&=U_{\mu}U_{\nu}T^{\mu\nu(2)}_{\text{linear}}\notag\\
               &=\tau^2\frac{\mathfrak{g}_{0}}{h^d}\Omega_{d-1}\Gamma(d+2)\zeta(d+1)T^{d+2}(-E^{\rho}_{(G)}\mathcal{D}_{\rho}\beta),\notag\\
        \Pi^{(2)}_{\text{linear}}&=\frac{1}{d}\Delta_{\mu\nu}T^{\mu\nu(2)}_{\text{linear}}\notag\\
               &=\tau^{2}\frac{\mathfrak{g}_{0}}{h^d}\Omega_{d-1}\Gamma(d+2)\zeta(d+1)T^{d+2}\left(-\frac{1}{d}E^{\rho}_{(G)}\mathcal{D}_{\rho}\beta\right),\notag\\
        j^{\mu(2)}_{\text{linear}}&={\Delta^{\mu}}_{\nu}N^{\nu(2)}_{\text{linear}}\notag\\
               &=\tau^{2}\frac{\mathfrak{g}_{0}}{h^d}\Omega_{d-1}\Gamma(d+1)\zeta(d)T^{d+1}\bigg[-\frac{1}{d}\dot{\beta}E^{\mu}_{(G)}+\frac{1}{d}B^{\mu\nu}_{(G)}\mathcal{D}_{\nu}\beta-\frac{1}{d^2}\psi E^{\mu}_{(G)}\label{transport coefficients2}\\
               &\quad-\frac{2}{d}\left(1-\frac{1}{d+2}\right)\Delta^{\mu \rho}E^{\sigma}_{(G)}\psi_{\rho\sigma}\bigg],\notag\\
        q^{\mu(2)}_{\text{linear}}&=-U_{\nu}{\Delta^{\mu}}_{\sigma}T^{\nu\sigma(2)}_{\text{linear}}\notag\\
               &=\tau^2\frac{\mathfrak{g}_{0}}{h^d}\Omega_{d-1}\Gamma(d+2)\zeta(d+1)T^{d+2}\bigg[-\frac{1}{d}\dot{\beta}E^{\mu}_{(G)}+\frac{1}{d}B^{\mu\nu}_{(G)}\mathcal{D}_{\nu}\beta-\frac{1}{d^2}\psi E^{\mu}_{(G)}\notag\\
               &\quad-\frac{2}{d}\left(1-\frac{1}{d+2}\right)\Delta^{\mu \rho}E^{\sigma}_{(G)}\psi_{\rho\sigma}\bigg],\notag\\
       \Pi^{\mu\nu(2)}_{\text{linear}}&={\Delta^{\mu}}_{\rho}{\Delta^{\nu}}_{\sigma}T^{\rho\sigma(2)}_{\text{linear}}-\frac{1}{d}\Delta^{\mu\nu}\Delta_{\rho\sigma}T^{\rho\sigma(2)}_{\text{linear}}\notag\\
               &=\tau^2\frac{\mathfrak{g}_{0}}{h^d}\Omega_{d-1}\Gamma(d+2)\zeta(d+1)T^{d+2}\left(\frac{4}{d(d+2)}B^{\rho(\mu}\Delta^{\nu)\sigma}\psi_{\rho\sigma}\right).\notag
\end{align}

\end{document}